\documentstyle[11pt,newpasp,twoside,psfig]{article}
\begin{document}

\title{SuperWASP: Wide Angle Search for Planets}
\author{R.~A. Street$^1$, D.~L. Pollacco$^1$, A. Fitzsimmons$^1$, F.~P.
Keenan$^1$, Keith Horne$^2$, S. Kane$^2$, A. Collier Cameron$^2$, T.~A.
Lister$^2$, C. Haswell$^3$, A.~J. Norton$^3$, B.~W. Jones$^3$, I. Skillen$^4$, 
S. Hodgkin$^5$, P. Wheatley$^6$, R. West$^6$, D. Brett$^6$}
\affil{$^1$APS Division, School of Physics, Queen's University of Belfast,
University Road, Belfast, BT7 1NN, Northern Ireland,\\
$^2$School of Physics and Astronomy, University of St. Andrews, North Haugh,
St. Andrews, Fife, KY16 9SS, Scotland, \\
$^3$Dept. of Physics and Astronomy, The Open University, Walton Hall, 
Milton Keynes, MK7 6AA, England, \\
$^4$Isaac Newton Group of Telescopes, Apartado de correos 321, E-38700 
Santa Cruz de la Palma, Tenerife, Spain, \\
$^5$Institute of Astronomy, University of Cambridge, Madingley Road,
Cambridge, CB3 0HA, England, \\
$^6$Dept. of Physics and Astronomy, University of Leicester, Leicester, 
LE1 7RH, England.  }

\begin{abstract} 

SuperWASP is a fully robotic, ultra-wide angle survey for planetary transits. 
Currently under construction, it will consist of 5 cameras, each monitoring a
$9.5^{\circ} \times 9.5^{\circ}$ field of view.  The {\em Torus} mount and
enclosure will be fully automated and linked to a built-in weather station. 
We aim to begin observations at the beginning of 2003.   

\end{abstract}

\section{Introduction}

The observations of the planetary transits of HD~209458 by Charbonneau et al.
(2000) and Henry et al. (2000) highlighted the role which can be played by
small aperture/ultra-wide field surveys in detecting transits by hot
Jupiter-type planets.  These surveys have the great advantage of requiring only
relatively inexpensive, off-the-shelf equipment, which is then dedicated to the
project.  The wide field of view allows many thousands of $\sim$7--13\,mag
stars to be photometrically monitored simultaneously.  As radial velocity
surveys indicate $\sim1$\% of these stars have hot Jupiters, we should, in
theory, be able to discover statistically significant numbers of these planets
within a reasonably short timescale.  Such a sample is necessary to answer
questions about the formation and evolution of these planets, and how it is
related to factors such as stellar metallicity, age, type, etc.  The magnitude
range matches that of the radial velocity programs, allowing detailed follow-up
observations.  

The {\em SuperWASP} project has developed from our experience in building and
operating the prototype {\em WASP0}.  This instrument is the subject of
separate papers by Kane et al. (this volume) and Street et al. (2000).  Having
proven that we can achieve the required high-precision photometry from this
single, manually-operated camera, we are now building a fully robotic,
multi-camera instrument supported by a custom-built mount.  {\em SuperWASP},
which is primarily funded by Queen's University, will initially be able to
monitor 5 separate fields simultaneously, with the potential for up to 10. 
This will provide precise photometry on $\sim$25,000 -- 50,000 stars at a time,
data which will form an important resource for bright star astronomy.  With
this in mind, we will be making use of the data to search for a number of
phenomena, including Near Earth Asteroids and optical transients in addition to
the primary goal of planet-hunting.  Here we present our science goals followed
by the equipment design, a discussion of the data we expect to gather and our
plans for its analysis and dissemination.  

\section{Science Goals}
\protect\label{sec:science}

\subsection{Planetary Transits}

From our {\em WASP0} dataset we estimate that {\em SuperWASP} will be able to
monitor between 5,000 and 10,000 stars per $9.5^{\circ} \times 9.5^{\circ}$
field.  We therefore expect to have simultaneous photometry for up to 50,000
stars.  We plan to observe a set of 5 selected fields continuously for 1--2
months at a time before moving on to the next set.   In order to estimate the
expected yield of transit detections from this data, we turn to the radial
velocity surveys which indicate that $\sim$1\% of Solar neighbourhood stars
harbour hot Jupiter companions.  Geometric arguments show that $\sim$10\% of
these planets should transit their parent stars.  Given the approximate numbers
of stars within {\em SuperWASP}'s magnitude range (7--13 mag), we can estimate
the expected number of planets detected per field.   Table~\ref{tab:yield}
compares the yield from {\em SuperWASP} to those from other similar projects,
e.g. VULCAN (Borucki et al. 2001).  

\begin{table*}
\centering
\caption{The expected yield of transit detections depends on the number of
stars monitored and hence the area of sky and the magnitude depth covered by
the survey.  {\em SuperWASP} will cover a $5 \times 9.5^{\circ} \times 
9.5^{\circ}$ field of view with 5 cameras, while a number of similar projects 
cover fields of $\sim6^{\circ} \times 6^{\circ}$.  }
\protect\label{tab:yield}
\vspace{5mm}
\begin{tabular}{lccccccc}
\hline
Area       	      	      	      	      	& 7 mag & 8 mag & 9 mag & 10 mag  & 11 mag  & 12 mag  & 13 mag \\
\hline
All sky       	      	      	      	      	& 0.5   & 2     & 8     & 32      & 128     & 512     & 2000   \\
$5 \times 9.5^{\circ} \times 9.5^{\circ}$	&     	& 0.02	& 0.1 	& 0.4 	  & 1.5     & 6.1     & 24     \\
$6^{\circ} \times 6^{\circ}$	      	      	&     	&     	& 0.01	& 0.04	  & 0.15    & 0.6     & 2.4    \\
\hline
\end{tabular}
\end{table*}

{\em SuperWASP} is designed to operate continuously round the year.  It
will therefore be able to monitor a maximum of 6 sets of 5, $9.5^{\circ}
\times 9.5^{\circ}$ fields per year; in reality this will be less due to
poor weather, technical problems, etc.  Table~\ref{tab:yield} indicates that
we can expect over one hundred transit detections per year.  This would
enable us to rapidly provide a large sample of planets for further 
analysis.  

\subsection{Near Earth Asteriods and Optical Transients} 

A dataset of continous, highly sampled photometry will be a extremely useful in a number
of areas of astronomy beyond planetary detection.  In particular, the dense sampling
will enable us to detect and closely examine the lightcurves of a range of transient
phenomena.  Consequentially, our reduction procedures aim to measure all objects present
in each frame.  Near Earth Asteriods and optical transients are of special interest to
our collaboration.  We expect that {\em SuperWASP} will set constraints on the number of
10 -- 100m asteroids in near-Earth space, and discover around 1/$\Box^{\circ}$/year
optical transient events, depending on $\gamma$-ray burst collimation models.  

\section{{\em SuperWASP}: The Design}
\protect\label{sec:equipment}

\begin{figure*}
\centering
\begin{tabular}{c}
\psfig{file=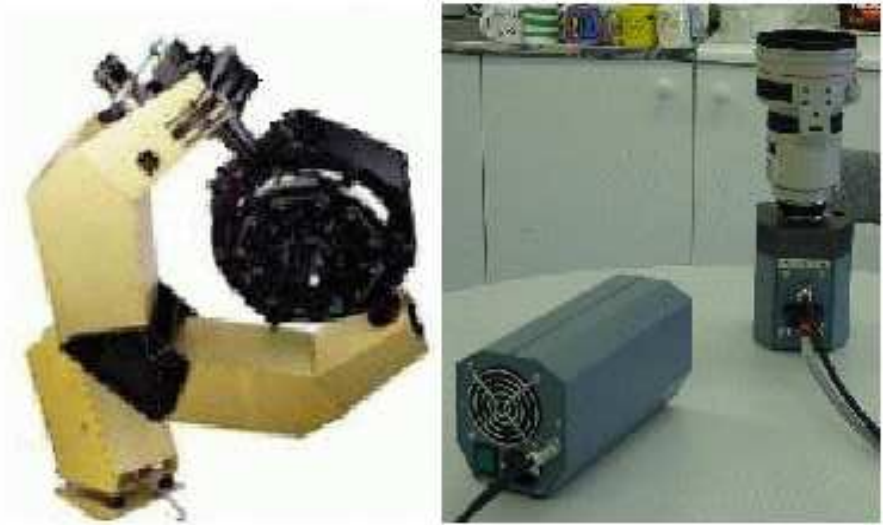,angle=270.0,width=12cm}
\end{tabular}
\caption{(Left) The robotic {\em Torus} fork mount which can operate with upto
10 cameras.  (Right) An example of a {\em SuperWASP} CCD camera
plus associated electronics.  }
\protect\label{fig:design}
\end{figure*}

Figure~\ref{fig:design} shows the {\em Torus} fork mount on which will be mounted the 5
CCD cameras (as shown on the right).  Each camera will consists of a $2048 \times 2048$
pixel thinned Marconi CCD and a Canon 200mm f/1.8 lens.  This combination will give each
camera a $9.5^{\circ} \times 9.5^{\circ}$ field of view while a pixel size of 13.5
$\mu$m means that the plate scale is expected to be 16.7 arcsec/pixel.  The cameras will
have an operating temperature of $-60^{\circ}$ C maintained by 3-stage Peltier cooling
mechanisms.  The readout time is expected to be 4s.  Five cameras will be mounted such
that each will be targeted individually, although the mounting is capable of supporting
up to 10 such cameras.   The {\em Torus} robotically-operated fork mount has a reported
pointing error over the whole sky of 30 arcsec and a tracking error of less than 0.01
arcsec per second.  A Global Positioning System receiver will give UTC with less than 1s
error.  The whole experiment will be housed inside an automated enclosure which will be
linked to a weather station.  The observatory will open robotically at dusk and observe
pre-determined target fields until automatic closure at dawn or if bad weather
intervenes.  

\section{{\em SuperWASP}: The Data}
\protect\label{sec:data}

With the advent of {\em SuperWASP} and projects like it, we are entering the
era of multi-terabyte astronomical datasets. From our {\em WASP0} experience,
we expect to take exposures of $\sim$30s, each of which will be 8.4MB in size. 
With a 4s-readout speed and perhaps $\sim8+$ hours of observations per
night, we expect to obtain at least  $\sim7$ GB of data per camera, every clear
night.  With 5 cameras, the expected data rate will be $\leq12.5$ TB per
year.   To handle such a large dataset, the data will be compressed
and written to DLT tape at the observatory.  These DLTs will be collected about
once a month and posted to Queen's University where backup copies will be made
and distributed to collaborators.  The reduction of the data will happen via an
automated pipeline as soon as the data is obtained, in order to avoid the
build-up of an insurmountable back-log.   This reduction will provide a
catalogue of photometric, positional and quality control data on every object
detected in every frame.  In this way, we will also be able to detect and
monitor transient events in great detail. This catalogue will then be mined and
analysed according to the aims of each project.  We aim to make this catalogue
publically available, probably via a web-based interface, as the data will
provide an invaluable resource for a large number of projects.  

\section{Summary}
\protect\label{sec:concs}

We present our plans for a multi-camera, robotic search for planetary
transits.  {\em SuperWASP} will have a greater field of view than other similar
projects, and a higher expected yield of planetary transits.  It will provide a
very large catalogue of densely sampled lightcurves of hundreds of thousands of
stars which could result in more than 100 planets being detected per year.  We
aim to make this data publically available.  {\em SuperWASP} is currently under
construction, and is on schedule to begin observations from La Palma in early
2003.

\end{document}